\documentclass[submission,copyright,creativecommons,noderivs]{eptcs}
\providecommand{\event}{LOCOCO 2010} 

\usepackage{breakurl}             
\usepackage[utf8]{inputenc}
\usepackage{algorithmic}
\usepackage{epsfig}
\usepackage{amssymb}
\usepackage{amsmath}
\usepackage{amsfonts}

\title{Comparison of PBO solvers in a dependency solving domain}
\author{Paulo Trezentos
\institute{ISCTE / Caixa Mágica}
\email{paulo.trezentos@iscte.pt}
}

\def\titlerunning{Comparison of PBO solvers in a dependency solving domain}
\def\authorrunning{P. Trezentos}
\begin{document}
\maketitle

\begin{abstract}
Linux package managers have to deal with dependencies and conflicts of
packages required to be installed by the user. As an NP-complete problem, 
this is a hard task to solve. In this context, several
approaches have been pursued. {\em Apt-pbo} is a package manager based on the apt
project that encodes the dependency solving problem as a pseudo-Boolean optimization 
(PBO) problem.
This paper compares different PBO
solvers and their effectiveness on solving the dependency solving problem.
\end{abstract}

\section{Introduction}

Software installation is the process of installing programs assuring that
specifically required software is pre-installed and that defined actions are taken
before or after the copy of the files into the file-system \cite{288441,787021}. Although this is a
common problem among Microsoft and Open Source Operating Systems (GNU/Linux,
BSD,...) \cite{1390640} we will focus on the later ones, since a progress in this
field would be applicable to all environments, including applications like
Eclipse or Firefox \cite{1595805}. 

The installation process comprises retrieving the package, solving the
software dependency tree, retrieving and installing the software dependencies
and finally installing the package and executing the associated install
scripts \cite{1595803}.

The dependency graph represents the software dependencies and
sub-dependencies needed for a package to work properly after installation \cite{Bix03}.
The restrictions imposed by the graph may have no solution (for instance, 
due to broken dependencies), only one solution, or several solutions. Criteria
such as the minimum number of packages or freshness can be defined to
rank the solutions in terms of their quality.
Finding a solution consists in defining the sub-set of packages that meets the dependency requirements. 
This process is called dependency solving. One approach to
dependency solving is to encode the problem as a pseudo-Boolean optimization (PBO) problem using
existing solvers for finding the optimal solutions. This approach is applied in {\em
  apt-pbo}, a meta-installer tool based on {\em apt} that will be described in this paper.

This paper is organized as follows. In section \ref{sec:back} we provide
background information about PBO. Section \ref{sec:sysoverview}
  depicts the {\em apt-pbo} tool and its architecture. Section
  \ref{sec:results} presents empirical results of experiments conducted with
  the several solvers. Finally, in section \ref{sec:conclusions} are presented
  the concluding remarks.

\section{Background}
\label{sec:back}


Pseudo-Boolean Optimization (PBO) is a special case of Integer Linear
Programming (ILP) where variables are Boolean. For this reason, it is 
often called 0-1 ILP.
This is the case of our package selection problem, where a package being
present in the final solution can be easily encoded as a Boolean variable being 
assigned value 0 or 1.

Pseudo-Boolean functions are a generalization of Boolean functions with a
mapping $\mathcal B^n = \{0,1\} \mapsto \mathbb{R} $
 \cite{barth95,772388}. Pseudo-Boolean functions in polynomial form are widely
used in optimization models in different areas like statistics, computer
science, VLSI design and operations research.

A PBO problem can be formally defined as follows \cite{barth95}:

\setlength{\arraycolsep}{0.0em}
\begin{eqnarray}
&\text{minimize } & \displaystyle\sum_{j \in \mathbb{N}} \  c_j \cdot x_j \\
&\text{subject to } & \displaystyle\sum_{j \in \mathbb{N}} \  a_{ij} l_j \geq
  b_i   \notag \\
& & x_j \in \{0,1\}, a_{ij},b_i,c_j \in \mathbb{N}_{0}^{+}, i \in M   \notag\\
& & M{}={}{1,...,m}   \notag
\end{eqnarray}
\setlength{\arraycolsep}{5pt}

where each $c_j$ is a non-negative integer cost associated with variable
$x_j$, $j \in \mathbb{N}$ and $a_{ij}$ denotes the coefficients of the literals
$l_j$ in the set of $m$ linear constraints, being a literal a Boolean variable 
or its negation.

Recent algorithms for solving the PBO problem integrate features from recent advances in
Boolean satisfiability (SAT) and classical branch and bound algorithms.

\section{System Overview}
\label{sec:sysoverview}

\subsection{Architecture}

{\em Apt} is a meta-installer widely used in Linux distributions. However, apt
solves dependencies in a very straightforward way and in a large number of
occurrences fails to deliver a solution. 

The {\em Apt-pbo} application \cite{aptpbo} belongs to a new generation of
meta-installers that not only are capable of finding a solution but are
flexible to allow the user to customize which solution fits best the needs. 

The architecture of {\em apt-pbo} has different hooks to integrate modules. This architecture allows exchange of modules. For example,
changing the PBO solver being used is an extremely easy task. 

In our tests, the overhead of the external calls is not significant since the
number of iterations is extremely low. 

\begin{figure}[htbp]
    \centering
    \includegraphics[width=10cm]{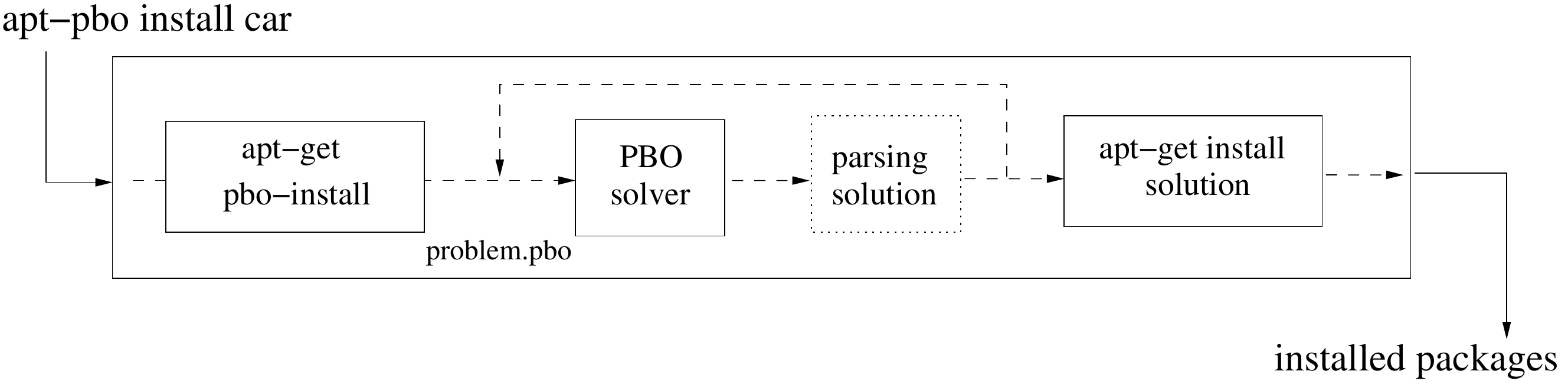}
  \caption{High level processing flow of {\em apt-pbo}}
  \label{fig:pipeline}
\end{figure}

Figure \ref{fig:pipeline} depicts a typical installation flow of apt-pbo.

The {\em apt-pbo} application is called with the operation {\em install} and the
desired package as arguments, which map the usage of {\em apt-get}.

The components of the figure have the following role:

\begin{itemize}
\item {\bf apt-get pbo-install}: we have a modified version of {\em apt-get} installation
software. Apt is one of the most used meta-installers and is adopted by 
Linux distributions like Debian, Ubuntu and Caixa M\'agica. The modifications
introduced by the author created a new method called {\em
  pbo-install}  which, given a specific package, calculates the  dependency tree
and writes the PBO encoding. The PBO encoding is composed of a set of PB-constraints and an
objective function.
\item {\bf PBO solver}: the {\em problem.pbo} formula is solved by 
  the PBO solver. We have used and tested different solvers as will be detailed in section
  \ref{sec:results}.
\item {\bf parsing solution}: apt-pbo has a module that parses the solver solution
  and, if necessary, establishes a new iteration with {\em apt-get
    pbo-install}. 
\item {\bf apt-get install solution}: when the final package set solution is reached,
  the user is asked for permission and the removal and installation of
  packages are performed using apt and dpkg / rpm.
\end{itemize}

\subsection{PBO encoding}
\label{sec:pboencoding}

As presented in the previous section, {apt-pbo pbo-install} encodes the
the problem as a Pseudo-Boolen Optimization.

This encoding has two parts: constraint and objective function definition-

{\bf Constraints definition}

In a pseudo-Boolean formula, variables have Boolean domains and constraints
are linear inequalities with integer coefficients. 

Encoding relations of the dependency tree as constraints is a straightforward task.
The following translations will be used:

\begin{itemize}
  \item {\em Installation}: $p_1$\footnote{For simplicity, the tuple representation of a package as $(p,2)$ will be now represented as $p_2$} is the package that we want to install: $p_1 \geq 1$. 
  \item {\em Dependency}: $p_1$ depends on $x_1$ should be represented as $x_1 - p_1  \geq
  0$. This means that installing $p_1$ implies installing $x_1$ as well, although
  $x_1$ may be installed without $p_1$. If $p_1$ also depends of $y_1$, we should add $x_1 - p_1  \geq 0$.
  \item {\em Multiple versions}: if a package $p_1$ requires the installation of a package $x$ 
    having different versions, for example $x_1$  and $x_2$, then we should encode the 
    requirement that installing package $p_1$ requires installing either package $x_1$ 
    or package $x_2$. Hence, such requirement may be encoded with constraint
    $x_1 + x_2 - p_1  \geq  0$.
  \item Conflicts: if a package has an explicit conflict with other package, for
  instance if $y_3$ conflicts with $x_1$, then this conflict is encoded as 
  $x_1 + y_3 \leq 1$. Remember that there is a conflict for each pair of 
  different packages corresponding to the same unit.
\end{itemize}

{\bf Objective Function Definition}

In the Objective Function we define what we plan to minimize. 

Two approaches might be followed: we minimize a single criterion (e.g. the
number of packages) ou multiple-criteria.

We will start by presenting single criterion.

\medskip

{\bf Minimizing Package Removal}

To minimize the number of removed packages, even if newer packages
exist, one should use the following {\em objective function}, where
$PI'_1..PI'_N$ is the set of packages already installed:

\begin{align*}
f_1(P) \text{  = min}  ( 1 - PI'_1) + ... + (1 - PI'_N)
\end{align*}

In order to minimize the objective function, the solver will try to set
variable $PI_i$ to 1 which will imply not removing installed applications.

{\bf Minimizing the Number of Installed Packages}

In this case, the total number of packages installed in the system is
to be minimized. Having $P1..PN$ as the new packages targeted to be
installed - either existent or new - the objective function will be:

\begin{align*}
f_2(P) = \text{\em min  P1 + ... + PN}
\end{align*}

\medskip

{\bf Maximizing the Freshness of Packages}

Consider $P1_1 .. P1_{k1}$ to be different software versions or releases of
package $P1$. Also, consider $v(P1_1)$ to be the normalized distance 
(a constant, for the purposes of the PBO problem) between the package $P1_1$ and the newest version present in repository 
$R$. Then the optimization function is:

\begin{align*}
f_3(P)=& \text{\em min } ( P1_1 * v(P1_1) + ... +  P1_{K1} * v(P1_{K1})) +\\
&( PZ_1 * v(PZ_1) + ... +  PZ_{KN} * v(PZ_{KN})) ...
\end{align*}

The value of $v(Pi_{Ki})$ is zero if the package is the newest in the repository.

\medskip

{\bf Multicriteria optimization}

However, in the real world installing a package follows multiple criteria and even if
one is more important than the others that can lead to non-desired solutions.

Trying to satisfy different criteria when finding the set of packages for a
software installation falls in the multicriteria decision making (MCDM) set of
problems \cite{Ehrgo00b}.

{\em Apt-pbo} integrates the different objective functions of the
previous section as a multiobjective problem (MOP):

\begin{align*}
 & \text{\em min } (f_1(P), f_2(P), f_3(P))
\end{align*}

with $P$ as the available packages and $f_1, f_2$ and $f_3$ as the existent objective functions.

The multiobjective problem is solved transforming it into a single objective
problem through {\em weighted sum scalarization}.

{\em Apt-pbo} uses the following coefficients, $\lambda$, representing the
overall utility for the user: Removal Cost - $W_{r}$ (weight given to the cost of a removal of
a package), Presence Cost - $W_{p}$ (weight given to the presence of a new or an
  already installed package) and Version Cost - $W_{v}$ (weight representing the cost of having an older
  version in the solution when a newer exists).

The objective function is then defined as:

\begin{align*}
& \text{\em min } (W_{r} \cdot f_1(P) + W_{p} \cdot f_2(P) + W_{v} \cdot f_3(P))
\end{align*}

\section{Experimental Results}
\label{sec:results}

We performed experiments on a large set of different repositories, packages
and systems hosted at O2H Lab cluster of 164 Xeon CPU cores\footnote{The infra-structure is integrated in the ADETTI / ISCTE centre
  of RNG Grid.} with Linux installed in Xen virtual system machines and inside a {\em chroot} environment. In what follows we report the results of this evaluation.

The goal of the experiments performed was to simulate the installation of
software in a Linux environment and test the different PBO solvers against the
same criteria.

A comparison of SAT and PBO solvers has been performed extensively through international
competitions and benchmarks \cite{satcompetition,pbocompetition,pb2009}.
Since the solving algorithm can benefit greatly from the structure of the
problem, it was considered important to evaluate different PBO solvers on solving 
this problem.
As mentioned in section \ref{sec:sysoverview}, {\em apt-pbo} is structured in a
modular form, thus allowing the replacement of one PBO solver by another 
compatible solver.

For testing purposes, four solvers were considered: 

\begin{itemize}
\item {\em minisat}+ \cite{minisat+-jsat06}: from the same authors of {\em minisat},
  a well known SAT solver, and actually based on {\em minisat},
  {\em minisat+} encodes PB-constraints into SAT.
\item {\em bsolo} \cite{heras2008}: bsolo is a PBO solver, which 
  was first designed to solve instances of the Unate and Binate
  Covering Problems (UCP/BCP) and later updated with pseudo-Boolean
  constraints support. 
\item {\em wbo} \cite{wbow2009}: from some of the same authors of bsolo, 
  participated in the PB'09 competition.
\item {\em opbdp} \cite{barth95}: an implementation in C++ of an implicit enumeration
  algorithm for solving PBO.  
\end{itemize}

Besides the solvers mentioned above, Pueblo \cite{Sheini06pueblo:a} was also 
considered but not included since the
only available version is dynamically linked and the libraries needed are old
and not available in the testing infra-structure. Nevertheless, an old Linux
system was installed (Debian Etch) and some ad-hoc tests were performed with
Pueblo. These tests revealed that Pueblo has a poor performance for this
specific type of problems and no further efforts to port Pueblo were made.

The tests consisted of 1,000 installation of packages over a Debian Lenny Linux
system. Two different scenarios were tested: ``conservative'' and ``aggressive''.

The weights in the objective function (section \ref{sec:pboencoding}) are the
same in both scenarios adopting a balanced configuration between updates and removals.

The difference are trhe active repositories. In the ``conservative'' scenario only Lenny repositories were active (main and
updates). In the ``aggressive'' the {\em Sid} (development
version) and {\em Backports} repositories were also present. Table \ref{tab:caract} summarizes
the differences between scenarios.  In fact, 12,000 more packages were present in
the ``aggressive'' scenario and more than the double of the total space accounted by
{\em apt-pbo} for mapping packages, dependencies and conflicts.

\begin{table}[h]
\centering
\caption{Characterization of packages - conservative and aggressive scenarios}
\begin{tabular}{|l||c|c|} \hline
Measures & Conservative & Agressive  \\ \hline\hline
Total package names & 30014 & 42007  \\ \hline\hline
Total distinct versions & 24100 & 51337 \\ \hline\hline
Total dependencies & 147085 & 326891  \\ \hline\hline
Total Provides mappings & 5146 & 10962  \\ \hline\hline
Total dependency version space & 602k & 1358k  \\ \hline\hline
Total space accounted for & 7284k & 14,9M   \\ 
\hline\end{tabular}
\label{tab:caract}
\end{table}

\subsection{Aggressive scenario}

Table \ref{tab:benchsolversag} summarizes the results of the evaluation
performed in the context of the aggressive scenario. 

As we can observe, both {\em wbo} and {\em bsolo} are able to solve all the instances but {\em wbo}
has a better performance (4.45 seconds on average per transaction). {\em Minisat+} comes in third place, not only with a lower
number of instances solved, 355, but also with a poorer performance, taking on
average more than two minutes to solve a problem. {\em wbo} has also a smaller standard
deviation than {\em bsolo}. The {\em average time} consists in the time, in average, per installation transaction.

\begin{table}[h]
\centering
\caption{PBO solvers benchmarking - Aggressive scenario }
\begin{tabular}{|l||c|c|c|c|} \hline
& bsolo & wbo & minisat+ & opbdp \\ \hline\hline
\# Solved &  1,000 & 1,000 & 355 & 47 \\ \hline
\# Timeouts & 0 & 0  & 645  & 953 \\ \hline
Average time & 00:07.79 & 00:04.45 & 02:30.16 & 07:16.49 \\ \hline
Standard deviation & 00:02.83 &  00:01.19 & 01:29.33 & 35:13.02  \\
\hline\end{tabular}
\label{tab:benchsolversag}
\end{table}

\begin{figure}[htbp]
    \centering
    \includegraphics[width=12cm]{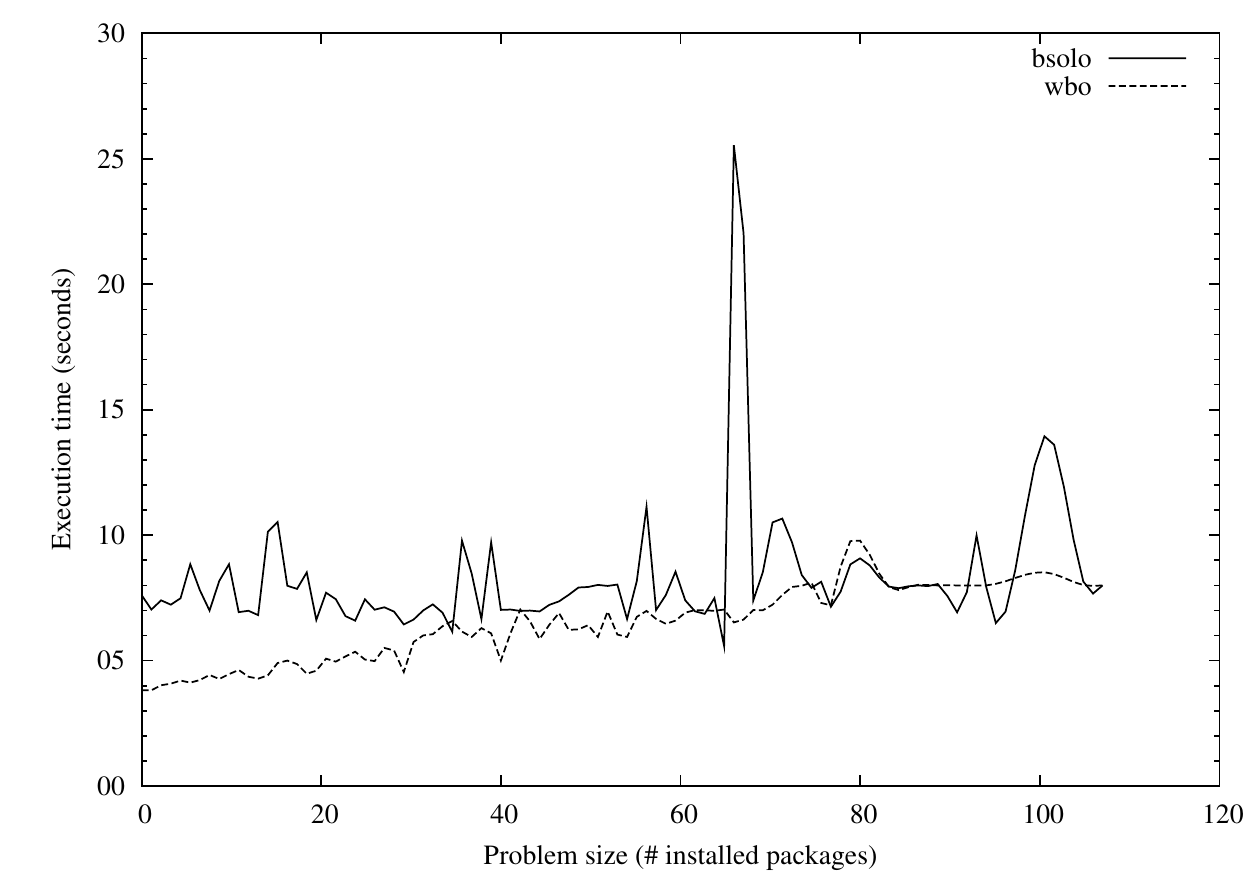}
  \caption{PBO solvers graph - Aggressive scenario}
  \label{fig:solversgraph}
\end{figure}

Figure \ref{fig:solversgraph} compares {\em wbo} and {\em bsolo} varying the
number of the installed packages per transaction.
There is a smooth growth by {\em wbo} and a
more unstable line of growth in a much more unpredictable fashion by {\em bsolo}.
 Since {\em minisat}+ and
{\em opbdp} had a significant number of timeouts, they were not included in
the graph. 

\subsection{Conservative scenario}

In the conservative scenario, development repositories are not active and
therefore there is a much more steady environment for dependency solving.

In this case, the four solvers were able to find the solutions before the
timeout of 150 seconds. In fact, on average they performed under 3 seconds
with the exception of {\em minisat+}. 

\begin{table}[h]
\centering
\caption{PBO solvers benchmarking - Conservative scenario }
\begin{tabular}{|l||c|c|c|c|} \hline
& wbo & bsolo & minisat+ & opbdp \\ \hline\hline
\# Solved & 1000  & 1000  & 1000 & 1000 \\ \hline
\# Timeouts & 0 & 0  & 0  & 0 \\ \hline
Average time & 00:02.6 & 00:02.62 & 00:06.22 & 00:02.55 \\ \hline
Standard deviation & 00:00.8 & 00:01.1  & 00:01.4 & 00:01.1 \\
\hline\end{tabular}
\label{tab:benchsolverscons}
\end{table}

\begin{figure}[htbp]
    \centering
    \includegraphics[width=12cm]{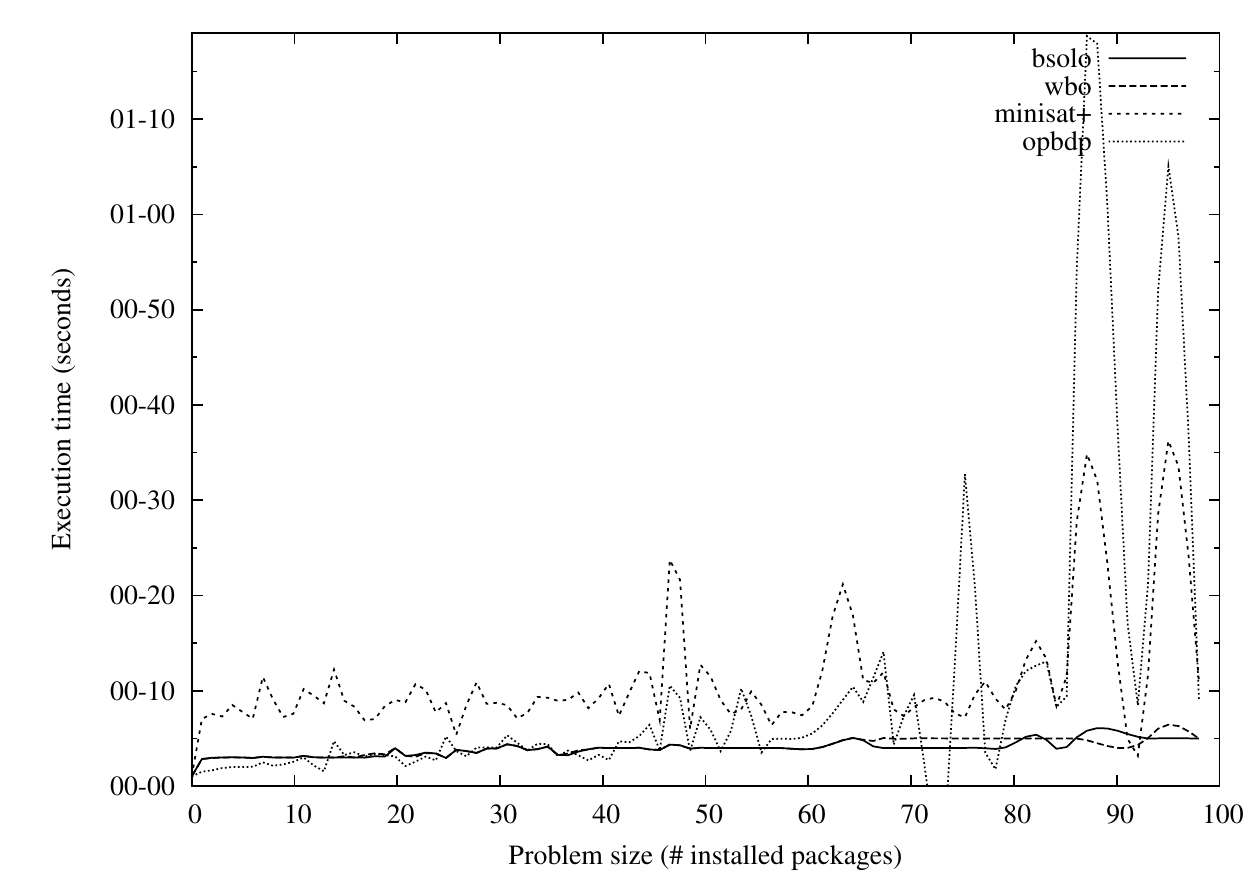}
  \caption{PBO solvers graph - Conservative scenario}
  \label{fig:solversgraph2}
\end{figure}

Figure \ref{fig:solversgraph2} depicts the size of the problem {\em vs}
time. Although on average opbdp performs better than minisat+, the figure shows that as
the size of the problem grows opbdp is more sensible to peaks and outliers.

\section{Related Work}
\label{sec:relatedwork}

The use of Boolean Satisfiability (SAT) \cite{HandbookOfSAT2009} for solving the dependency problem
has first been proposed in the context of the EDOS FP6 
project \cite{EdosAse06,wp2d2} which had impact in other research efforts \cite{le-berre:sat-for-dependency}. An alternative formulation using 
constraint programming techniques has been described in \cite{mouthuy06}, including the 
use of different heuristics for improving the quality of the solution found.


\section{Conclusions}
\label{sec:conclusions}

The PBO solvers evaluated follow different theoretical approaches and therefore are
expected to have different results. However, some results of the tests performed are
interesting to recall: {\em wbo} is the solver that performed better in both
scenarios and with a more stable behaviour. {\em bsolo} has also
interesting results in both scenarios.

Although {\em wbo} is the solver with better time results, there are other
aspects to take in account: {\em minisat+} is open source and can be
enhanced to address more difficult problems as the presented ones in the aggressive
scenario. Being open source is a critical point to a Linux distribution that might adopt
such a tool.

Future work will consist in analysing, jointly with the authors of the PBO tools,
possible enhancements of the tools as a result of this evaluation. Another
direction for future work is to study the possibility of the solvers 
returning a non-optimal solution when the timeout is reached.

Finally, this article can be extended to study other solvers such as SCIP \cite{Achterberg2004TR} and
boolean optimization engines such as SAT4JPB \cite{berre10} or MsUnCore \cite{msuncore09}.

\subsection{Acknowledgments}

Partially supported by the European Community's 7th Framework Programme (FP7/2007-2013), grant agreement n${}^\circ$214898.


\bibliographystyle{eptcs} 

\end{document}